\documentclass[reprint,3p]{elsarticle}

\usepackage{subcaption}
\usepackage{color}

\usepackage{amsmath,amssymb}

\usepackage{lineno,hyperref}

\begin{document}

\begin{frontmatter}

\title{Towards implementation of Cellular Automata in Microbial Fuel Cells}


%
%

\author[label1]{Michail-Antisthenis Tsompanas\corref{cor1}}
\address[label1]{Unconventional Computing Centre, University of the West of England, Bristol BS16 1QY, UK\fnref{label1}}
\cortext[cor1]{Corresponding author}
\ead{antisthenis.tsompanas@uwe.ac.uk}

\author[label1]{Andrew Adamatzky}

\author[label4]{Georgios Ch. Sirakoulis}
\address[label4]{Department of Electrical and Computer Engineering, Democritus University of Thrace, Xanthi 67100, Greece\fnref{label4}}

\author[label2]{John Greenman}

\author[label2]{Ioannis Ieropoulos}
\address[label2]{Bristol BioEnergy Centre, University of the West of England, Bristol BS16 1QY, UK\fnref{label2}}
\ead{ioannis.ieropoulos@brl.ac.uk}


\begin{abstract}
The Microbial Fuel Cell (MFC) is a bio-electrochemical transducer converting waste products into electricity using microbial communities. Cellular Automaton (CA) is a uniform array of finite-state machines that update their states in discrete time depending on states of their closest neighbors by the same rule. Arrays of MFCs could, in principle, act as massive-parallel computing devices with local connectivity between elementary processors. We provide a theoretical design of such a parallel processor by implementing CA in MFCs. We have chosen Conway's Game of Life as the `benchmark' CA because this is the most popular CA which also exhibits an enormously rich spectrum of patterns. Each cell of the Game of Life CA is realized using two MFCs. The MFCs are linked electrically and hydraulically. The model is verified via simulation of an electrical circuit demonstrating equivalent behaviors. The design is a first step towards future implementations of fully autonomous biological computing devices with massive parallelism. The energy independence of such devices counteracts their somewhat slow transitions --- compared to silicon circuitry --- between the different states during computation.
\end{abstract}


\end{frontmatter}

\section*{Introduction}
Microbial Fuel Cells (MFCs) are devices that produce electricity from waste-water by utilizing microbial metabolic oxidation processes \cite{ieropoulos2013energy}. A MFC consists of a proton exchange membrane (PEM), an anode (negative half-cell) and a cathode (positive half-cell). Electricity is generated as a by-product of microbial metabolism, which results in electrons flowing from the bacterial cells onto the anode electrode, and then from anode to the cathode electrode via an external electrical circuit; this produces a flow of electrical current. Positively charged ions that are products of the oxidation reactions, such as protons, also flow out of the bacterial cells, due to electron-neutrality and diffuse to the cathode through the PEM. Electrons, cations and the cathodic oxidizing agent of choice (e.g. oxygen) recombine to complete the reaction and close the circuit. Despite the fact that MFC technology has been the subject of research for at least three decades, real-world implementation or commercialization is still limited \cite{cheng2011electricity,santoro2015cathode,ortiz2016study,mardanpour2012single,ledezma2013mfc,chouler2016towards}. 

In addition to the utilization of MFCs in waste-water treatment and power production, their usage in other fields was suggested, like sensory applications \cite{kumlanghan2007microbial,abrevaya2015analytical}. Nonetheless, an application that is rarely studied is using configurations based on MFCs that realize computing units \cite{greenman2006microbial,greenman2006perfusion}. More specifically, the first approach \cite{greenman2006microbial} was to reproduce conventional binary logic gates using MFCs in order to examine the abilities of such a system, with an ultimate goal of constructing non-silicon multi-valued logic processing units being envisaged by the same authors in \cite{greenman2006microbial}. In that study, designs of hydraulic and electrical interconnections are suggested for building three basic logic gates (AND, OR and NOT) that can be combined to assemble universal gates, hence circuits capable of universal computation.  

Apart from basic logic gates, more complicated computational abilities were exhibited with the appropriate interconnection of a small number of MFCs, namely a simplified Pavlovian learning model \cite{greenman2006perfusion}. In this study, the symbiotic mix of natural biological cells, such as the anodophiles, and artificial systems, like electrodes, actuators, pumps and chemical solutions, were used to simulate a learning cycle. In detail, two signals representing the smell of food --- that activates production of saliva --- and hearing the sound of a bell --- that does not activate independently the production of saliva --- were associated, in order for the sound of the bell to trigger the production of saliva by itself, i.e. automatically.

Despite the seemingly simplistic, binary computational tasks performed in both of the aforementioned studies \cite{greenman2006microbial,greenman2006perfusion}, the authors make notice of the high number of states that could be realized by MFC devices, in order to enable complex computing. In this respect, a configuration of MFCs is proposed here to mimic the computation dynamics of Cellular Automata (CAs). A novel development of MFCs, which includes the introduction of additional electrodes, acting as poise/bias points --- or `pins' --- has been recently proposed [Patent filing number: GB1501570.4]. The invention introduces the principle of electrochemical redox (reduction--oxidation) bias via a third and/or fourth electrode, when an external power supply, or another MFC is connected to this third or fourth pin and to the working anode or cathode electrode, respectively. This is, by default, an unconventional means of connection, since the potential difference (voltage) of the external MFC (also known as `driver') can bias the redox potential difference (voltage) of the anode or cathode half-cell, depending of course on how the connection is made. Such a system is naturally subject to polarization and is therefore limited to a time constant ($t$), based on materials, voltage levels and oxidation/reduction states. Consequently, the system is an ideal platform for pulse-width-modulation techniques. For the purposes of the current study, a third electrode is used to achieve poise and, thus, have two MFCs behaving like a CA cell. 

CAs can be considered an idealization of a physical system in which space and time are discrete, and the physical quantities take only a finite set of values \cite{chopard2009cellular,mizas2008reconstruction}. A CA comprises identical cells in a regular grid that are characterized by their state. The state of each cell is updated by a uniform local rule depending on the states of the cells in its vicinity. CAs can conceptually be identified as \textit{general} and \textit{simple} \cite{sipper1995quasi}. The term \textit{general} is referring to the fact that CAs can promote universal computation and that the state and local updating rules of the cells are not limited to specific regulations. Moreover, the term \textit{simple} is justified by the plain outline of CAs --- cells are characterized by basic states with local interactions --- compared with other computing machines. Finally, CAs can be considered as one of the most favorite candidates of the future computational architectures tackling the bottleneck of the von Neumann architecture when referring to the co-existence of computing and memory units in the same simple unit, or the CA cell. 

Specifically, a well known CA model, namely Conway's Game of Life (GoL) is studied, which encapsulates the ability of universal computation and construction \cite{adamatzky2010game}. However, the realization of GoL here is not limiting the capabilities of possible configurations consisted of hydraulically and electrically linked MFCs. In fact, based on the local activity \cite{chua1998cnn} exhibited by MFCs, any CA local rule can be implemented in similar configurations. In addition to the advantage of the energy independence, along with water purification of the proposed computational scheme, when compared with other transducers of renewable energy sources, MFCs integrate both energy extracting mechanisms and computational units. Moreover, the ongoing miniaturization of MFCs \cite{chouler2016towards} will enable the production of smaller biological processing units. Finally, the amount of physicochemical parameters that can be externally manipulated and affect the performance and, thus, the outputs of the biofilms in MFCs is enormous \cite{greenman2006perfusion}. This fact can justify the utilization of MFCs for more complex computational schemes than the ones suggested up to this date.

\section*{Game of Life}

A Cellular Automaton (CA) consists of a regular grid of cells. Each cell takes $k$ different states, where $k > 2$, but not at once. The grid can be $n$-dimensional ($n \geq 1$). The evolution of the cells takes place at discrete points in time. That means that the state of each cell in the grid changes only at discrete moments of time, namely at time steps $t$. The time step $t = 0$ is usually considered as the initial step and therefore no changes at the state of the cells occur.

For each cell, a set of cells called its neighborhood (usually including the cell itself) is defined relative to the specified cell. Regarding the two dimensional CA, the two most common types of neighborhood that are mainly considered are:

\begin{itemize} 
\item \textit{von Neumann} neighborhood, that consists of the central cell, whose condition is to be updated, and the four cells located to the north, south, east and west of the central cell

\item \textit{Moore} neighborhood, that consists of the same cells with the von Neumann neighborhood together with the four other adjacent cells of the central cell (the northwester, northeaster, south-east and south west cells).
\end{itemize}

The evolution of the cells demands the definition of a cell state, the neighboring cells as well as the local transition function:

\begin{itemize}
\item The local internal state of each cell of the CA:
\begin{equation}
C ( \vec{r},t) = \{ C_{1}( \vec{r},t), C_{2}( \vec{r},t),..., C_{m}( \vec{r},t)\}
\label{theory_state}
\end{equation}
\noindent at time step $t=0,1,2,...$ is described by a set of variables associated with each position $\vec{r}$ of the grid.

\item The local transition function is defined as:
\begin{equation}
R=\{R_{1},R_{2},...,R_{m}\}
\label{theory_function}
\end{equation}
\noindent and determines the evolution during time of the internal state  of each cell according to the following equation:
\begin{equation}
C_{j}( \vec{r},t+1) =R_{j} \left( C( \vec{r},t), C( \vec{r}+\vec{\delta}_{q},t),..., C( \vec{r}+\vec{\delta}_{m},t) \right)
\label{theory_evolution}
\end{equation}
\noindent where $\vec{r}+\vec{\delta}_{k}$ designate the cells which belong to a given neighborhood of cell $\vec{r}$.
\end{itemize}

The state of cell $\vec{r}$, at time step ($t+1$), is computed according to $R$. $R$ is a function of the state of this cell at time step ($t$) and the states of the cells in its neighborhood at time step ($t$). In the above definition, the function $R$ is identical for all sites and it is applied simultaneously to each of them, leading to synchronous dynamics. It is important to notice that the rule is homogeneous, i.e. it does not depend explicitly on the cell position $\vec{r}$. However, spatial inhomogeneities can be introduced by having some cells' states $C_{j}(\vec{r})$ systematically at a fixed value, i.e. 1, in some given locations of the lattice, to mark particular cells for which a different rule applies. Furthermore, the new state at time $t+1$ is only a function of the previous state at time $t$. It is sometimes necessary to have a longer memory and introduce a dependence on the states at times $t-1$, $t-2$, …, $t-k$. Such a situation is already included in the definition, if one keeps a copy of the previous state in the current state.

Conway's Game of Life (GoL) is a two-dimensional CA with binary states \cite{conway1970game} that had significantly contributed to the extensive attention CA theory has gained. The neighborhood considered is \textit{Moore} neighborhood and the two states that each cell can adopt is \textit{alive} and \textit{dead} (or `1' and `0', respectively). The local transition rule uses the states of all nine cells in the neighborhood, during the directly preceding time step, to determine the new state of the central cell in the neighborhood. More specifically, the following transitions between states can occur:

\begin{enumerate}

\item When a cell is \textit{dead} at time \textit{t} and precisely three of the eight neighbors are \textit{alive}, the cell adopts the state \textit{alive} at time \textit{t}+1.

\item When a cell is \textit{alive} at time \textit{t} and none, one or more than three of the eight neighbors are \textit{alive}, the cell adopts the state \textit{dead} at time \textit{t}+1.

\end{enumerate}

Note that if none of the two aforementioned cases are true, the local rule dictates that the cell retains its previous state. Assuming $i$ and $j$ the dimension indexes that establish the location of each cell in the grid and $t$ the current time step, the transition rule can be expressed as:

\begin{equation}
C_{i,j}^{t+1}= \begin{cases}
        0, & \text{if }  (\sum_{k=-1,l=-1}^{k=1,l=1} C_{i+k,j+l})-C_{i,j} \leq 1 \text{ or} \\
        & (\sum_{k=-1,l=-1}^{k=1,l=1} C_{i+k,j+l})-C_{i,j} \geq 4 \\
        1, & \text{if } (\sum_{k=-1,l=-1}^{k=1,l=1} C_{i+k,j+l})-C_{i,j} = 3\\
        C_{i,j}^{t} & \text{else } 
        \end{cases}
\label{golrule}
\end{equation}

Note that \textit{Moore} neighborhood of cell $C_{i,j}$ is consisted of cells $C_{i+1,j}$, $C_{i,j+1}$, $C_{i-1,j}$, $C_{i,j-1}$, $C_{i+1,j+1}$, $C_{i+1,j-1}$, $C_{i-1,j+1}$, $C_{i-1,j-1}$ and the cell itself.

It is suggested that the inherent complexity of GoL is due to the fact that its transition rule is non-monotonic and nonlinear \cite{adamatzky2010game,chua1998cnn,Rendell2002}. Moreover, the aforementioned rule is characterized as an outer totalistic rule, given that it only accounts for the value of the central cell during the last time step and on the sum of values of cells in the outer \textit{Moore} neighborhood.

A continuous version of GoL, namely a discrete-time, continuous spatial automaton with the same behavior as GoL has been also proposed \cite{maclennan1990continuous}. In continuous spatial automata, the cells and their states form a continuum. A similar behavior to GoL in a continuous field can be realized with a local rule implemented by a non-monotonic function of the population density in the neighborhood. That is assuming the function will be graphically represented in two dimensions, namely the input or the population density in $x$-axis and the output or the next state of the cell in $y$-axis. Keeping in mind the rules of the original GoL, the continuous local rule function has to be continuously increasing in the space $(0,m)$ and continuously decreasing in the space $(m,1)$, for instance an inverted parabola; where $m$ is a value in space $(0,1)$, the minimum value of population density is $0$, while the maximum value is $1$. Note that value $m=3/8$ provides the closest analogy to the standard, binary GoL.

Another study \cite{adachi2004game} has proposed a CA model with a local rule based on a continuously valued expression with three parameters. That model matches GoL when one of these parameters, namely one defined as temperature $T$, is approximately zero and the other two have appropriate values. The upper limits of the temperature parameter were investigated, where formations of GoL, like gliders, start to decay. Nonetheless, it is suggested that for higher values of $T$ parameter, the model's behavior is increasingly biased towards chaos \cite{adamatzky2010game}. 

Furthermore, the realization of GoL by Cellular Neural Networks (CNNs) have been suggested \cite{274337}. In that study, a two layer single-step CNN template, a multi-step three layer discrete time CNN template with threshold sigmoid and a multi-step and a multi-step piecewise-linear discrete time CNN template have been presented.

\section*{Proposed implementation}

\begin{figure}[tb]
				\centering
				\begin{subfigure}[b]{0.45\textwidth}
					\includegraphics[width=\textwidth]{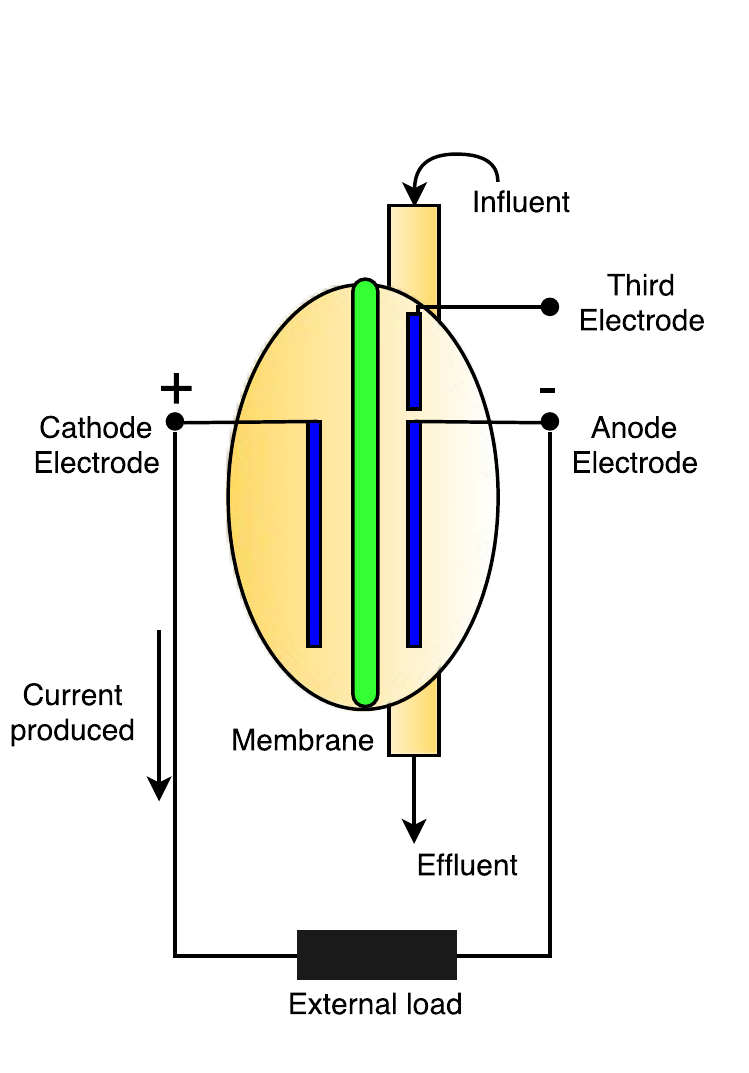}
					\caption{}
					\label{fig1a}
        \end{subfigure}
        \begin{subfigure}[b]{0.4\textwidth}
					\includegraphics[width=\textwidth]{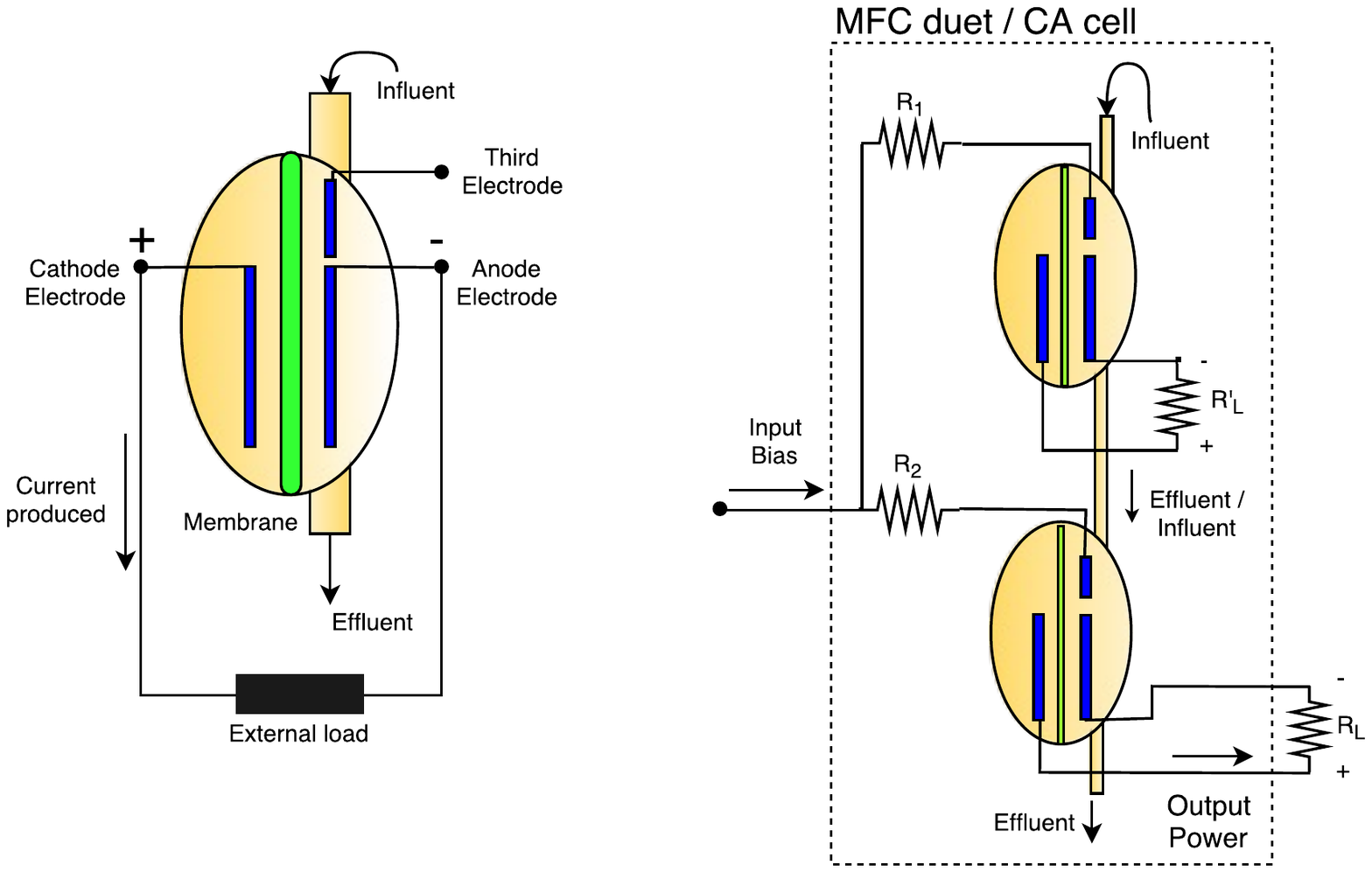}
					\caption{}
					\label{fig1b}
        \end{subfigure}
\caption{{\bf Schematic of the configuration of MFCs implementing a GoL CA cell.} A: Inputs and outputs of one MFC. B: The configuration of a MFC duet.}
\label{fig1}
\end{figure}

Drawing inspiration from an implementation using MFCs to build logic gates \cite{greenman2006microbial}, we propose a new implementation to execute a popular example of CA, namely the aforementioned GoL. The instance of GoL is selected here as it is an excellent example of the fact that complex behaviors emerge from a trivial local interaction of simple agents. Nonetheless, the application of the GoL rules to realize functions that can be translated as global computations \cite{adamatzky2010game}, can define the limitations of the proposed configuration. The functionality of a MFC is affected by the voltage applied on the third electrode (see Fig. \ref{fig1}A) of the device, which introduces an electrochemical redox bias.

The scheme realizing one cell of GoL CA consists of two MFCs, one primary and one secondary. The two MFCs are both hydraulically and electrically connected as shown in Fig. \ref{fig1}B. The secondary MFC is fed by a main/initial source of fuel, while the primary is fed by the effluent of the second. Both are operating under continuous flow conditions. MFCs stacked in cascades have been proven to produce higher power and current densities when their position is higher up the cascade, thus fed directly by the fuel source, than the ones placed lower downstream \cite{ledezma2013mfc,walter2015microbial}. Consequently, setting a fuel source that provides a balanced but limited substrate concentration, will enable only one of the MFCs to function.

The selection of which of the two MFCs will be functional is a result of their electrical interconnection. Both MFCs are equipped with a third electrode, independently. These electrodes are connected, via different resistances, with a single pin representing the electrical input of the CA cell implementation. Note that the different values of resistance separate the operation of the MFC GoL cell into three regions depending on the input voltage applied as explained below. The output power of the primary MFC is used to describe the state of the CA cell, i.e. its output. 

\begin{figure}[!tb]
				\centering
				\begin{subfigure}[b]{0.4\textwidth}
					\includegraphics[width=\textwidth]{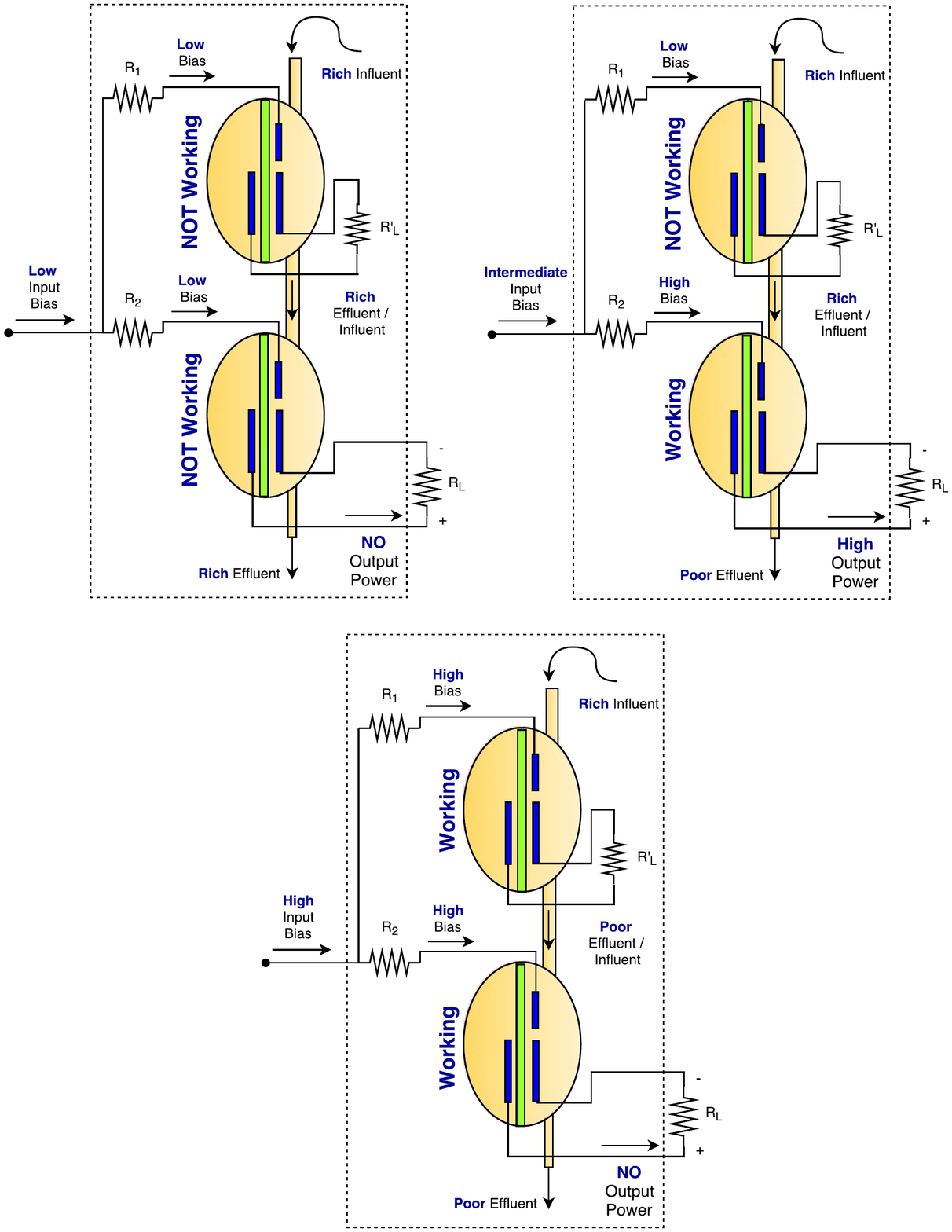}
					\caption{}
					\label{fig2a}
        \end{subfigure}
        \begin{subfigure}[b]{0.4\textwidth}
					\includegraphics[width=\textwidth]{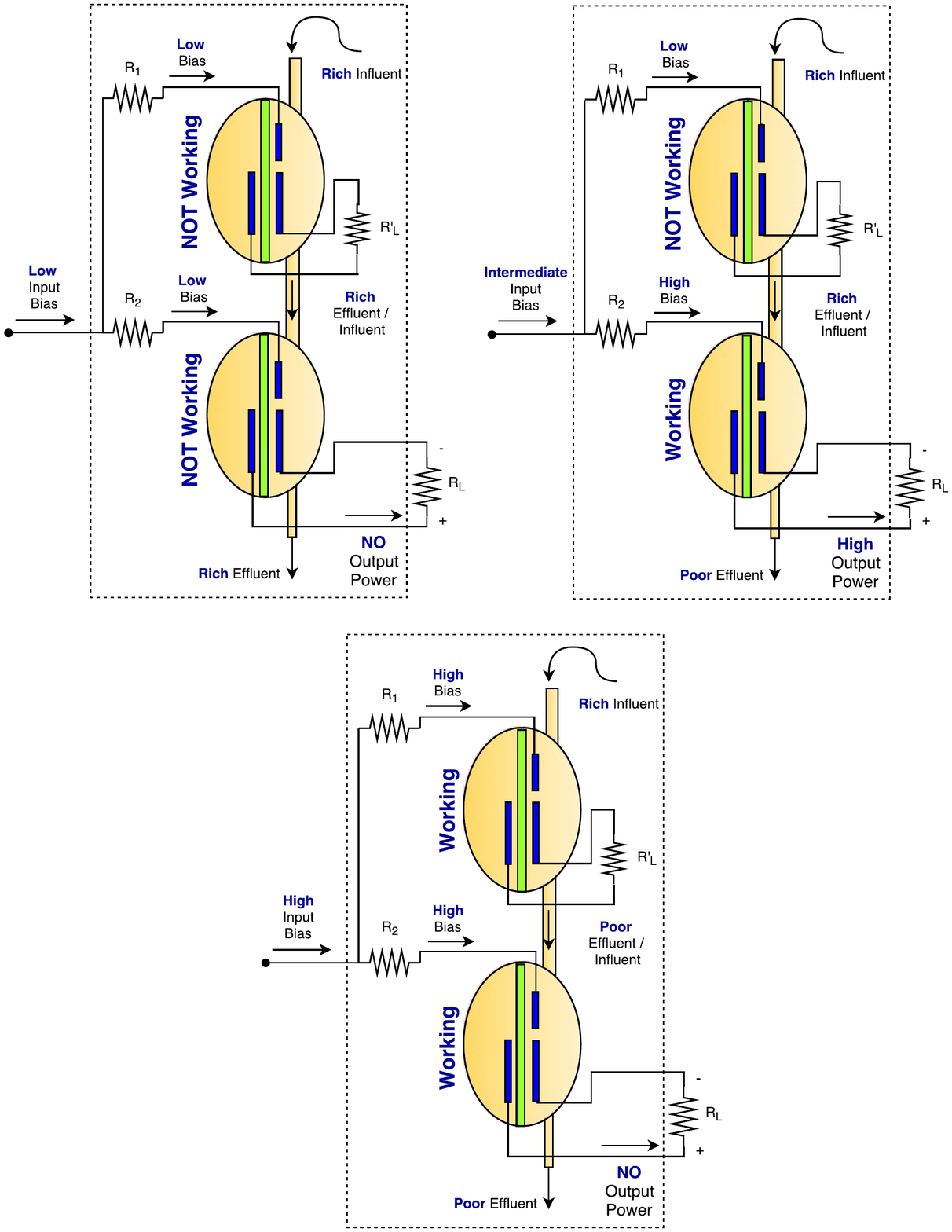}
					\caption{}
					\label{fig2b}
        \end{subfigure} \\
				\begin{subfigure}[b]{0.4\textwidth}
					\includegraphics[width=\textwidth]{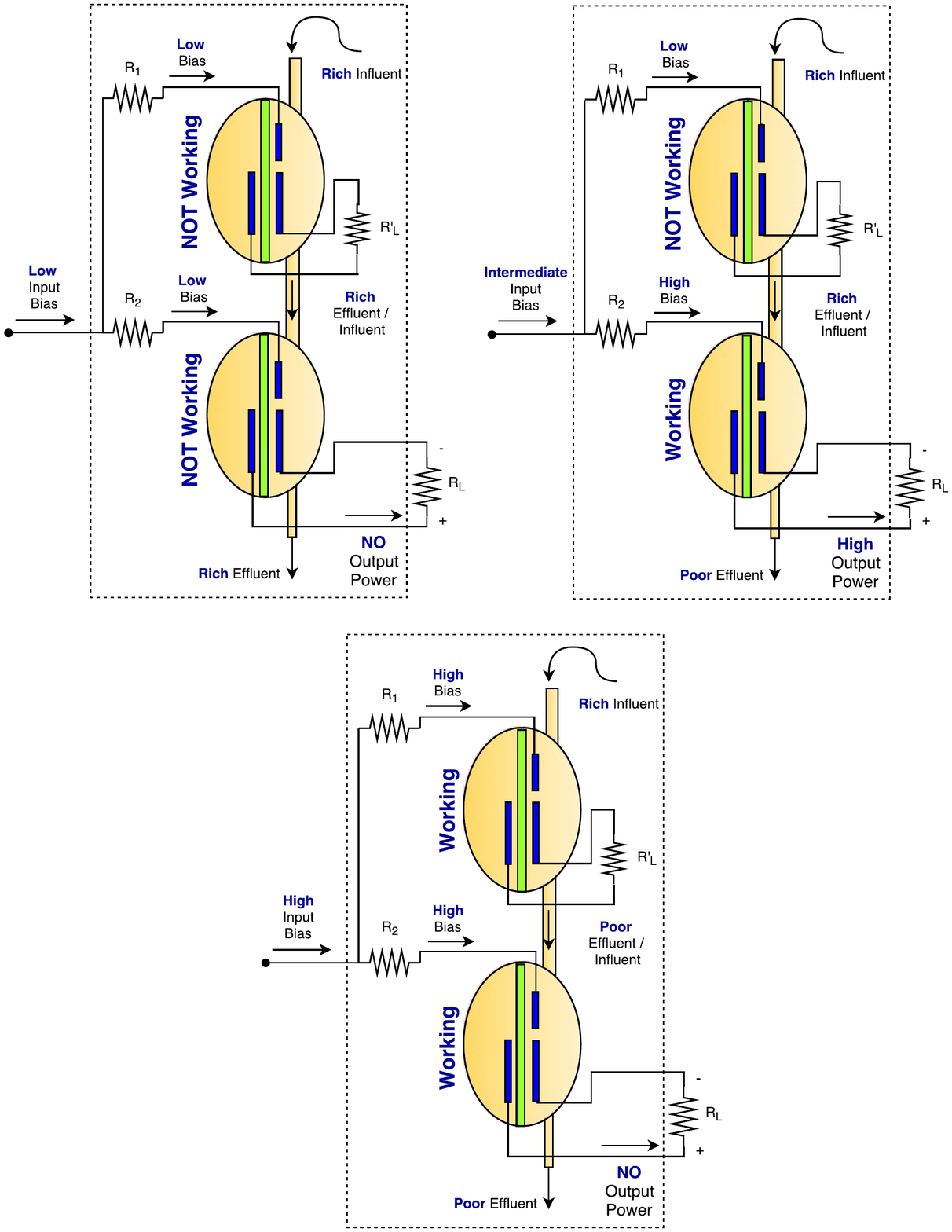}
					\caption{}
					\label{fig2c}
        \end{subfigure}
\caption{{\bf The three operation regions of the MFC-based scheme.} A: Low input bias resulting state `0'. B: Intermediate input bias resulting state `1'. C: High input bias resulting state `0'.}
\label{fig2}
\end{figure}

The secondary MFC will act as a control unit, through the hydraulic link, for the primary one. The main fuel source is considered to provide a solution with limited carbon-energy, so that only one of the two MFCs will be able to fully process this by its anodophilic biofilm in order to produce electricity. When an appropriate bias is applied to the input port of the CA cell, thus, on the third electrode of the secondary MFC, the processes in the biofilm of the anode of the secondary MFC will be activated. Consequently, the biofilm will be utilizing the nutrients from the source, resulting in an effluent depleted from carbon-energy, which is used as the influent for the primary MFC. This means that the primary MFC will be unable to produce electrical energy and its power output will be low, hence the state of the proposed CA cell will be `0' (see Fig. \ref{fig2}C).

On the other hand, when the bias applied to the third electrode of both MFCs is low enough (or zero), neither of them will be activated. Consequently, the power produced by the primary MFC will be low and the state of the proposed CA cell will be `0' (see Fig. \ref{fig2}A).

Finally, an intermediate value of bias applied to the CA cell input, or both MFCs' third electrodes, will allow the primary MFC to produce electricity. Provided a design with appropriately chosen values of the resistances connected to the third electrodes, the voltage drop over the resistance connected to the secondary MFC will be sufficiently large not to enable its activation and, thus, the depletion of carbon-energy in its effluent. Note that the values of the resistances should satisfy $R_1 > R_2$, in order to set voltage drops that will activate only the primary MFC. The voltage drop over the resistance connected to the primary MFC will not be large enough, but given that the influent will be carbon-energy replete (i.e. rich in metabolites), it will result in the production of electrical current (state `1') (see Fig. \ref{fig2}B).

The power output ($P$) of each CA cell for the three regions of operation is expressed in relation to the input bias applied ($V$) as in Eq. \ref{eq1}. Note that the time variant is included in the equation, to conform to CA terminology. The time variant has an actual real analogy with the transition response for the establishment of a new steady-state, resulted in changes in the conditions applied to the biofilm component. That transition response is identified as approximately four minutes \cite{greenman2006microbial}. 

\begin{equation} \label{eq1}
P_{out}^{t+1} =  \begin{cases}
        P_{low}, & \text{for }  V_{in}^{t} \leq V_{thr\_low}\\
        P_{high}, & \text{for } V_{thr\_low}\leq V_{in}^{t}\leq V_{thr\_high}\\
        P_{low}, & \text{for } V_{thr\_high}\leq V_{in}^{t}
        \end{cases}
\end{equation}

\begin{figure}[!tb]
\centering
\includegraphics[width=0.5\textwidth]{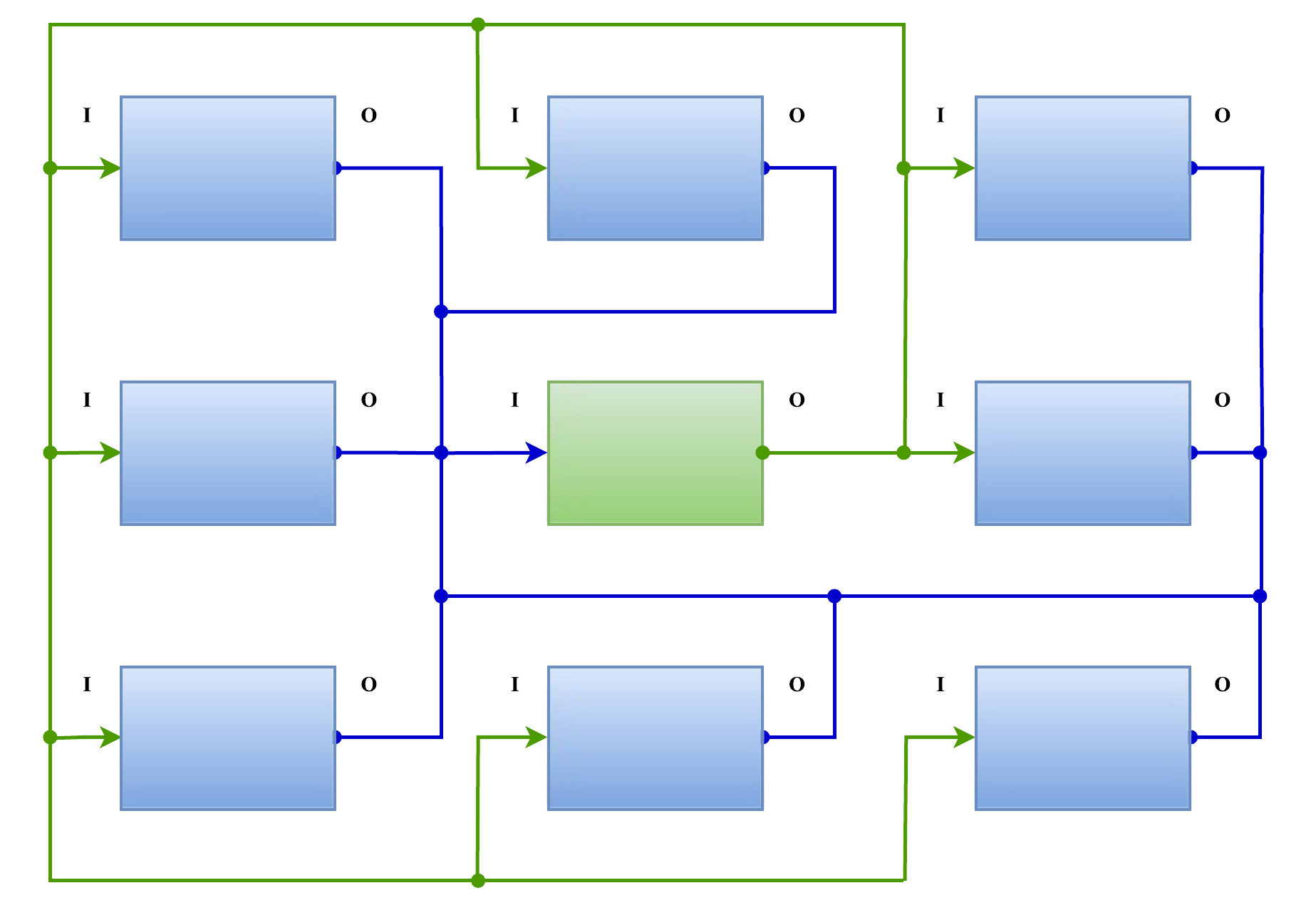}
\caption{{\bf Electrical interconnection of neighboring CA cells.}}
\label{fig3}
\end{figure}

Each CA cell, realized by a duet of MFCs, is connected with its eight neighbors, as illustrated in Fig. \ref{fig3}, to form \textit{Moore} neighborhood which is used in GoL. The current produced by each primary MFC is used to convey the information about central cell's state to all of its neighbors. 

\section*{Electrical Circuit Equivalent}
\begin{figure}[!tb]
\centering
				\begin{subfigure}[b]{0.45\textwidth}
					\includegraphics[width=\textwidth]{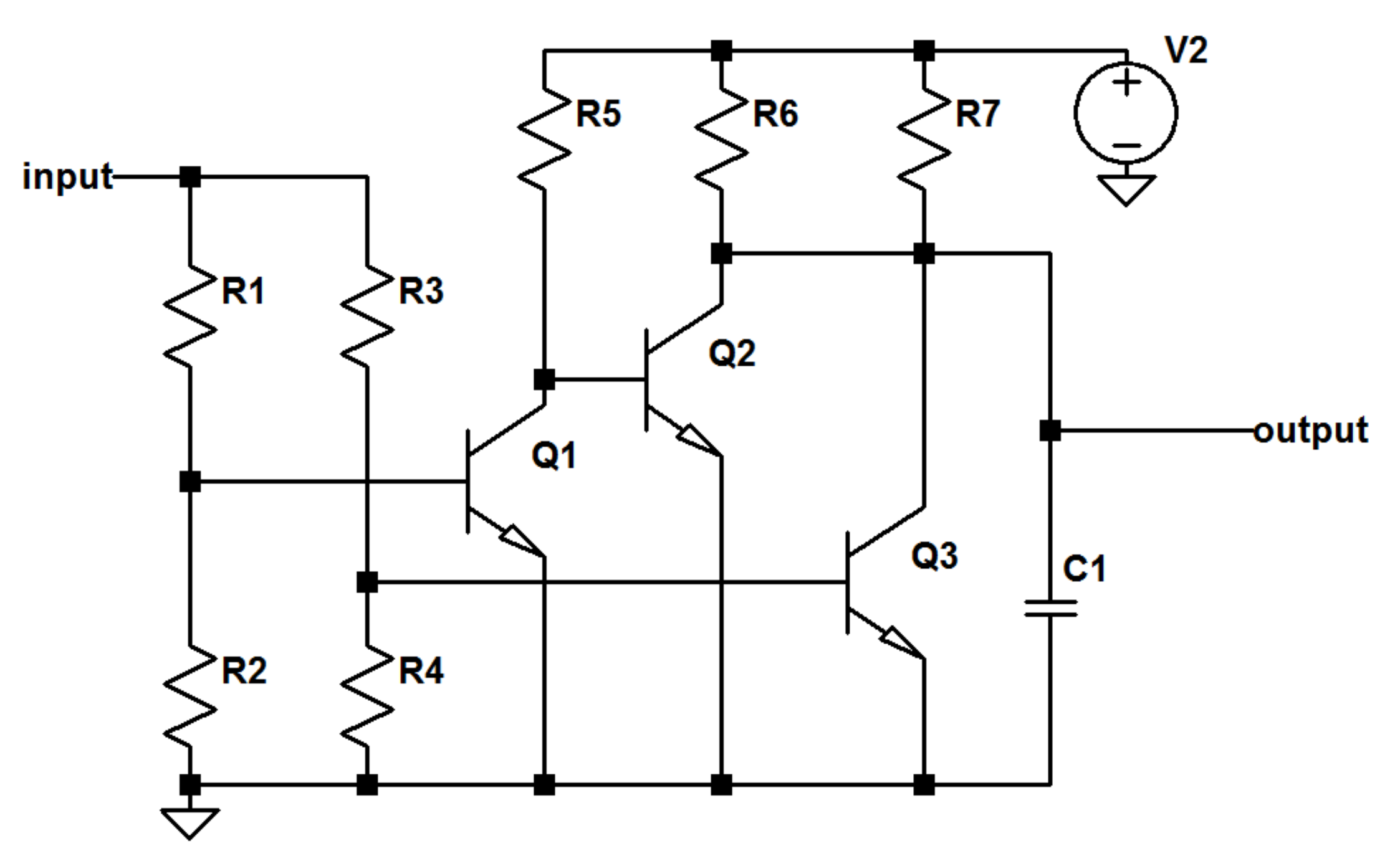}
					\caption{}
					\label{fig4a}
        \end{subfigure}
        \begin{subfigure}[b]{0.4\textwidth}
					\includegraphics[width=\textwidth]{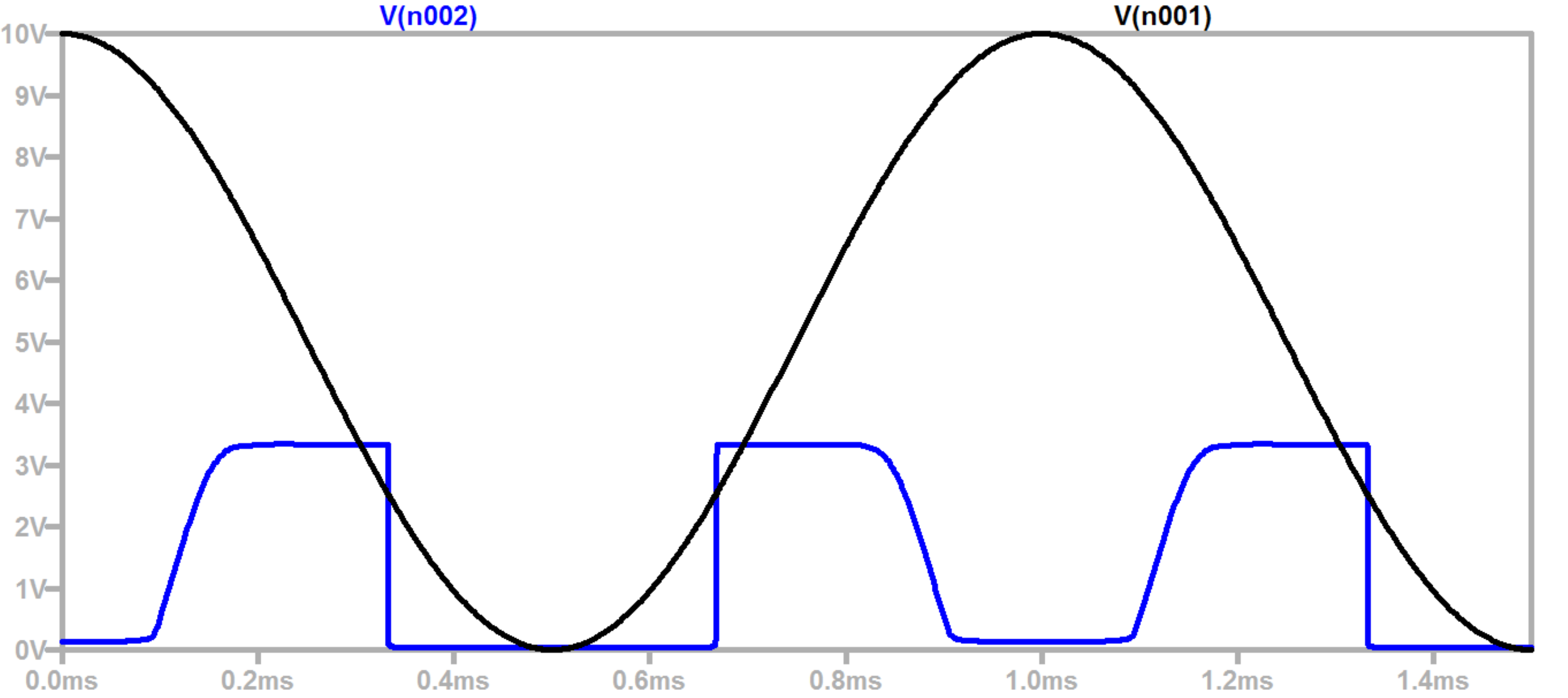}
					\caption{}
					\label{fig4b}
        \end{subfigure}
\caption{{\bf Equivalent electrical circuit to the proposed MFC configuration.} A: Schematic of the equivalent circuit. B: Output of equivalent circuit.}
\label{fig4}
\end{figure}

An electrical circuit that has an equivalent behavior to the MFC configuration presented in the previous Section and, thus, implements the GoL cell state transition rule, is presented in Fig \ref{fig4}A. However, there are significant differences. Firstly, the MFC logic can combine hydraulic and electrical links, thus, it can accommodate more complex computations with the same amount of basic building units. As mentioned previously, there is an enormous amount of physicochemical parameters that affect biofilms in MFCs. As a result, a MFC --- that is utilized as a computing unit --- except from the electrical pins, has also hydraulic inputs that affect its functionality. Note that there is no possible connection of two transistors that will act as an electrical counterpart of the proposed MFC duet configuration and, thus, in Fig. \ref{fig4}A three transistors are used. Moreover, MFC schemes do not require an external power supply, but a fuel source; a fuel that is inexpensive and abundant. Consequently, the proposed computing configurations instead of consuming energy, they are able to produce electrical energy, the level of which will depend on the requirements of the desired task; in other words, a more complicated computational task, requiring a higher number of MFCs, will naturally generate more power. In contrast, the transition response is faster in the electrical circuit.

The circuit consists of three transistors in total. Transistors \texttt{Q1}-\texttt{Q2} ensure a high output when the input voltage is higher than a lower threshold (i.e. 2V as depicted in Fig. \ref{fig4}B), while \texttt{Q3} results in a high output when the input voltage is lower than a high threshold (i.e. 7V as depicted in Fig. \ref{fig4}B). Also, the interconnection of collectors of transistors \texttt{Q2} and \texttt{Q3} forms a hard-wired {\sc and} gate. The voltage output (blue line) of the circuit correlated with a sinusoid voltage input (black line) is illustrated in Fig. \ref{fig4}B. The behavior of the circuit imitates the behavior of a continuous GoL cell \cite{maclennan1990continuous}.

\begin{figure}[!tbp]
\centering
				\begin{subfigure}[b]{0.4\textwidth}
					\includegraphics[width=\textwidth]{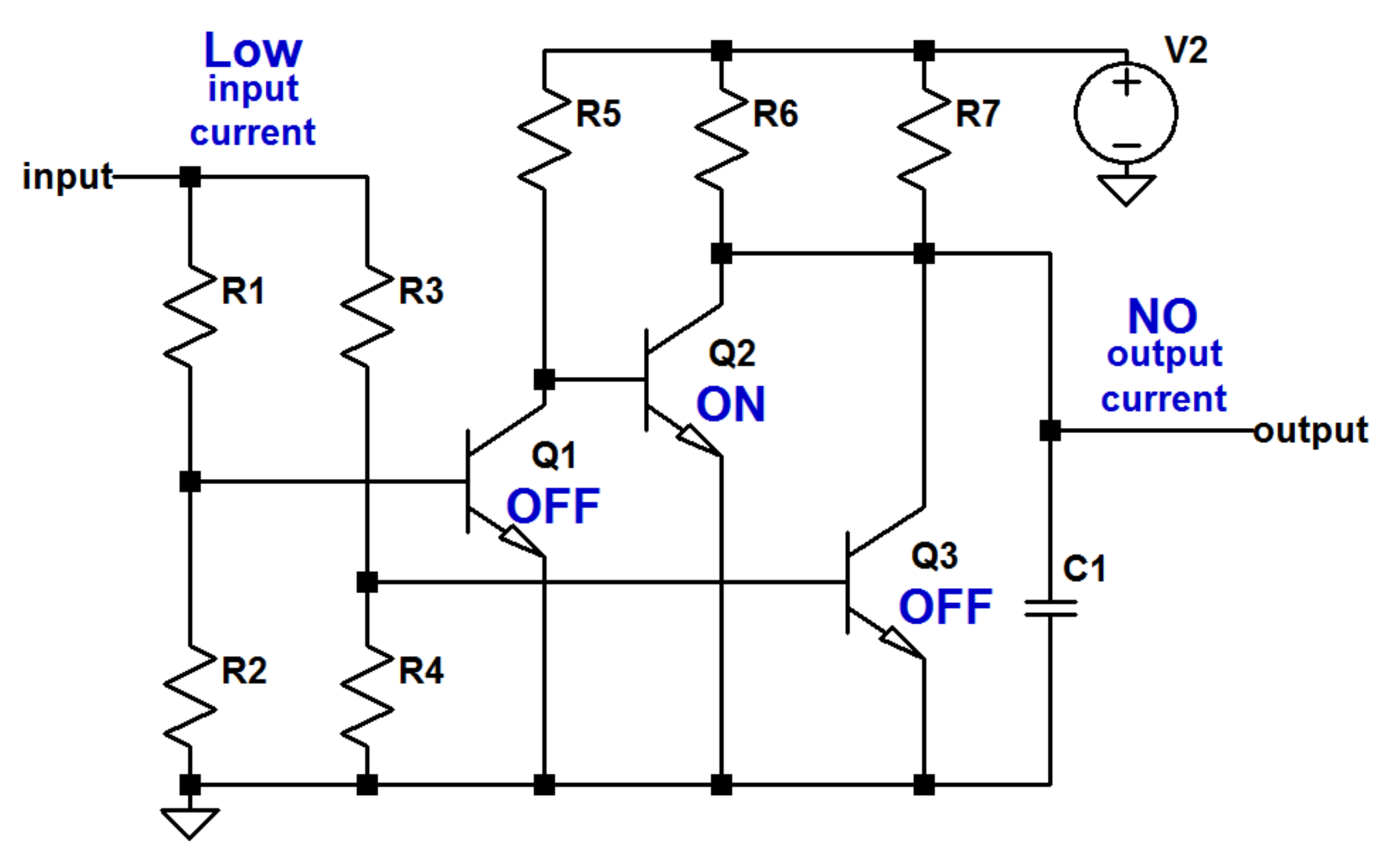}
					\caption{}
					\label{fig5a}
        \end{subfigure}
        \begin{subfigure}[b]{0.4\textwidth}
					\includegraphics[width=\textwidth]{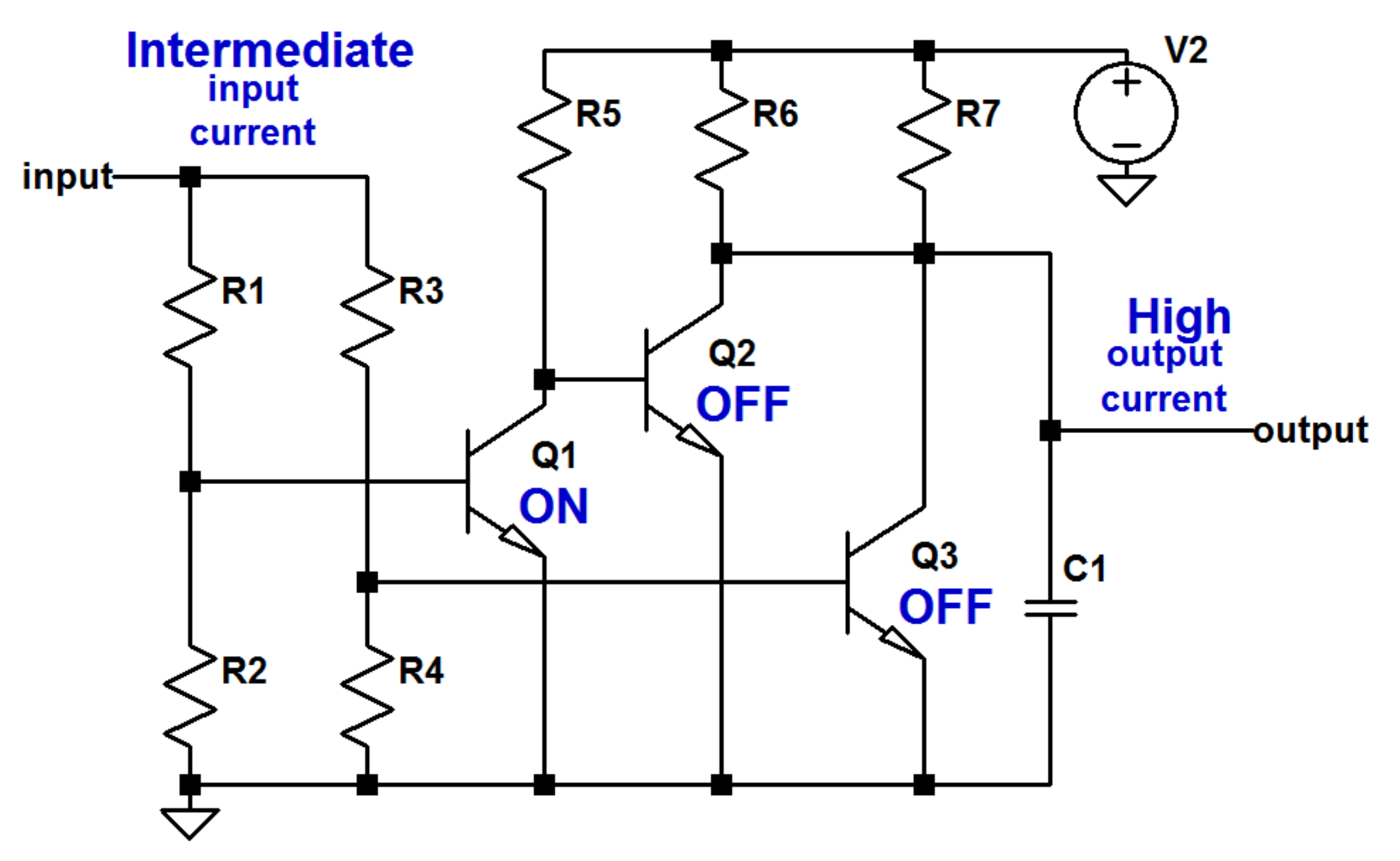}
					\caption{}
					\label{fig5b}
        \end{subfigure} \\
				\begin{subfigure}[b]{0.4\textwidth}
					\includegraphics[width=\textwidth]{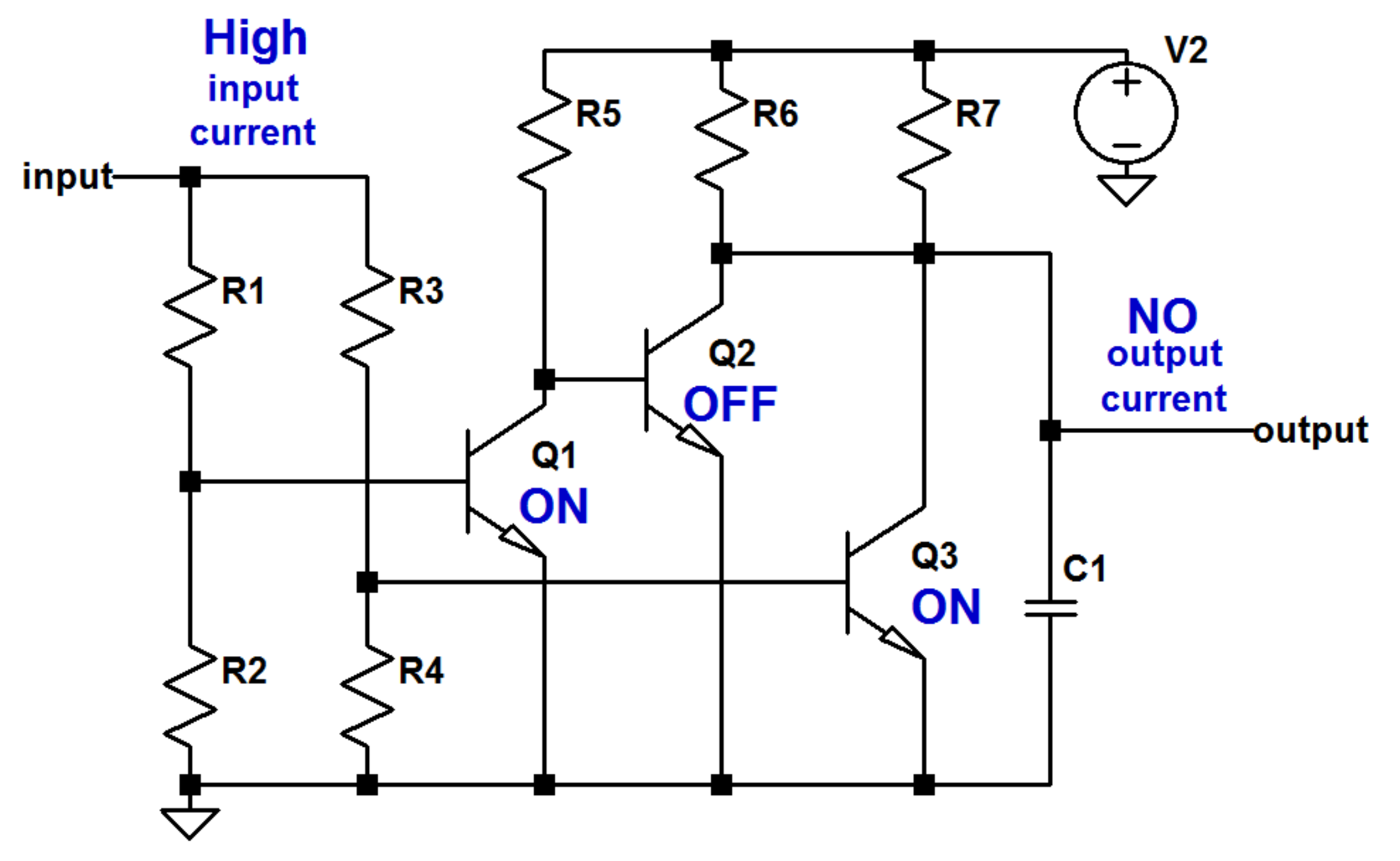}
					\caption{}
					\label{fig5c}
        \end{subfigure}
\caption{{\bf The three operation regions of the equivalent circuit.} A: Low current input resulting state `0'. B: Intermediate current input resulting state `1'. C: High input current resulting state `0'.}
\label{fig5}
\end{figure}

The functionality of the equivalent circuit for the three different states is presented in Fig. \ref{fig5} and can be compared with the functionality of the MFC configuration presented in Fig. \ref{fig2}.

\begin{figure}[!tbp]
\centering
\includegraphics[width=0.9\textwidth]{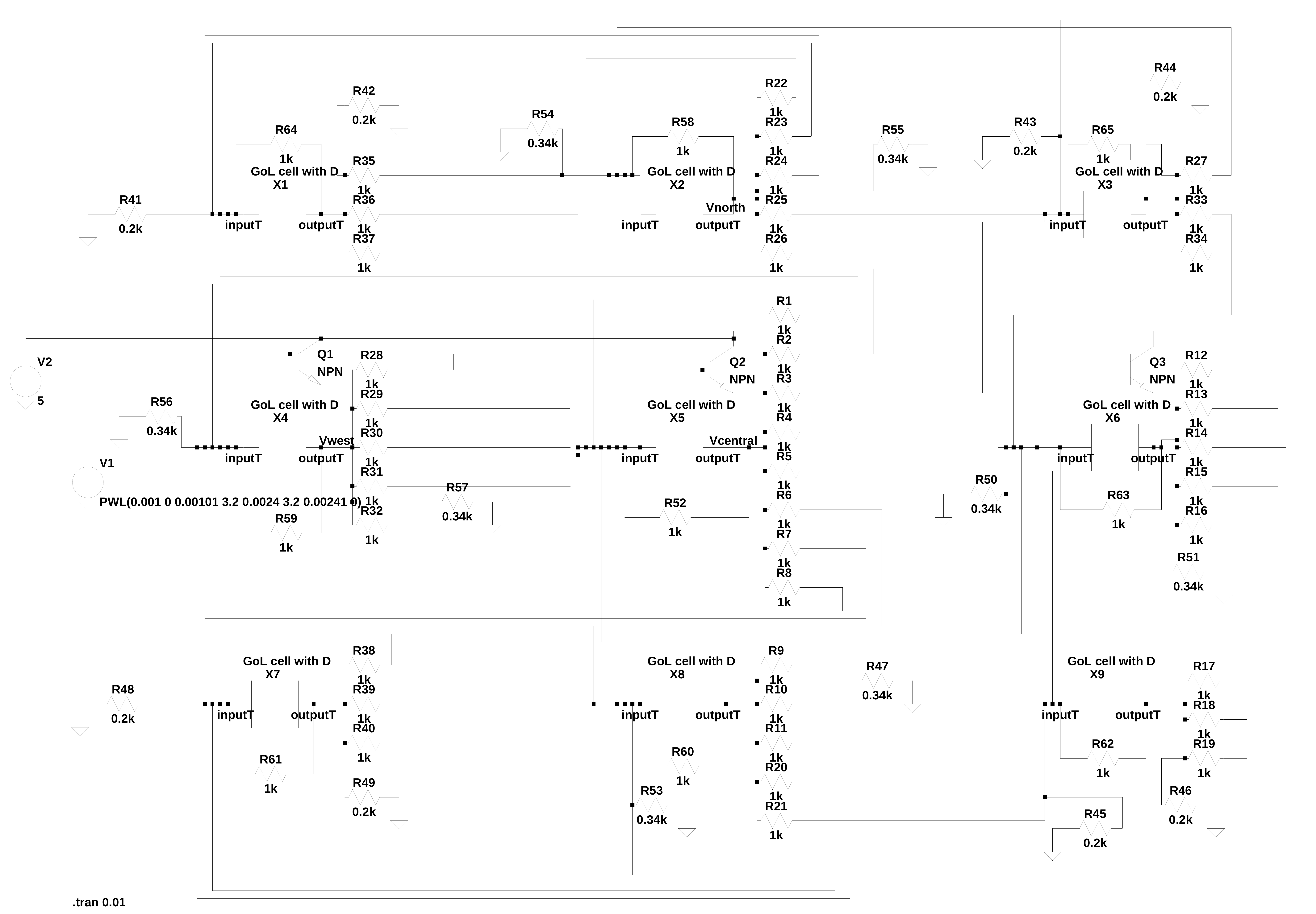}
\caption{{\bf A $3 \times 3$ grid of equivalent circuit cells.}}
\label{fig6}
\end{figure}

\begin{figure}[!tbp]
\centering
\includegraphics[width=0.7\textwidth]{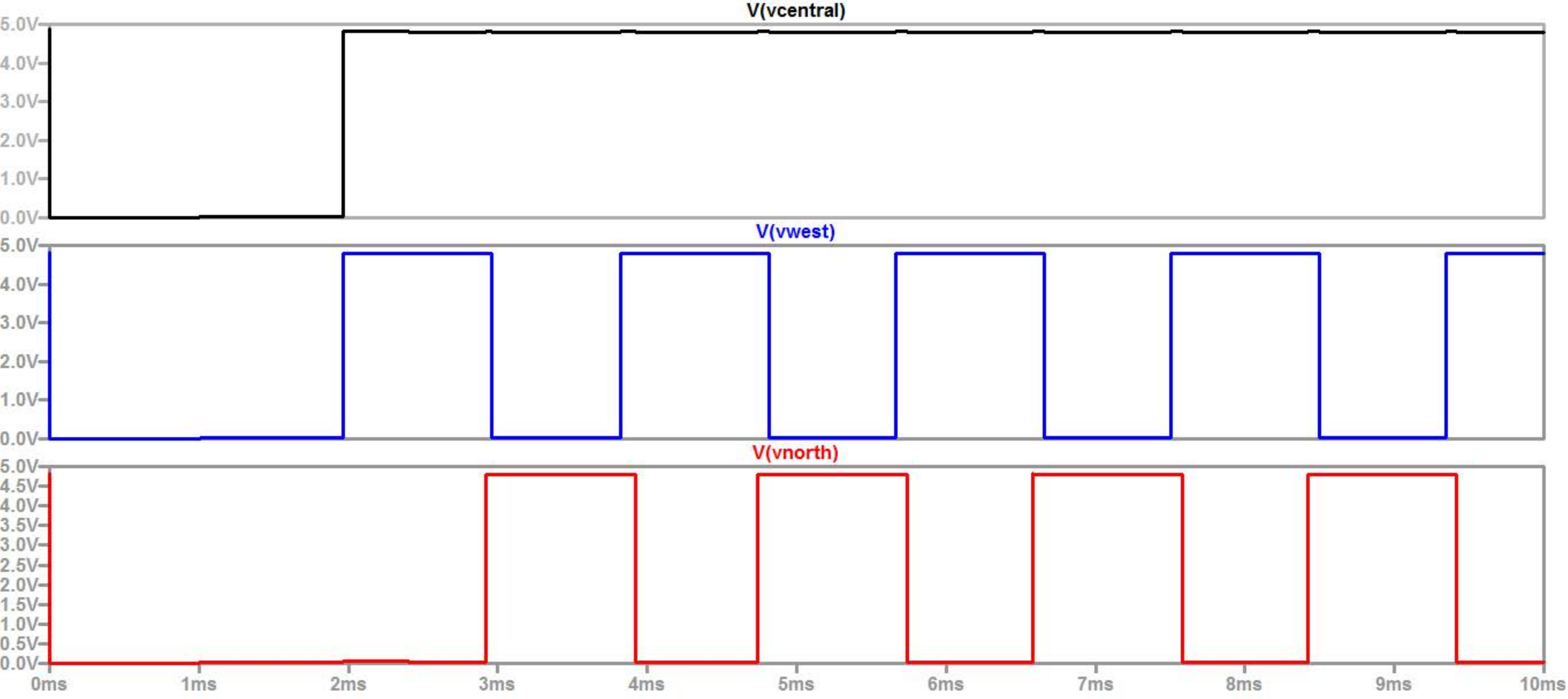}
\caption{{\bf The results of the grid initialized to oscillate as a GoL blinker.}}
\label{fig7}
\end{figure}

In order to illustrate the functionality of the equivalent circuit under the rules of GoL, a grid of  $3 \times 3$ cells is designed using LTspice software as shown in Fig. \ref{fig6}. Despite the fact that the length of the grid can be characterized as small, this grid is used for demonstration reasons, thus its simplicity enhances readability and in detail comprehension of the proposed electronic circuits. Nevertheless, designing larger grids, i.e. of $n \times n$ cells, is a trivial and effortless procedure, due to the well known prominent and inherent characteristics of CA, like local interconnections, simplicity, uniformity and area utilization. The grid is initialized using transistors \texttt{Q1}, \texttt{Q2} and \texttt{Q3} to set the inputs of cells \texttt{X4}, \texttt{X5} and \texttt{X6} to the voltage needed to trigger the `1' state in the next time step. Note in the results depicted in Fig. \ref{fig7} that the outputs of the central cell (\texttt{X5}) and the west cell (\texttt{X4}) are high (state `1') at $t=2ms$. Also, note that the state of the central cell remains `1', while the states of the west (\texttt{X4}) and the north (\texttt{X2}) cells are oscillating between `1' and `0' states, whereas they are not both in the same state for any time step. Each cell in the grid presented in Fig. \ref{fig6} is comprised of the circuit depicted in Fig. \ref{fig4}A and a circuit adding some time delay between its input and its output. The time delay circuit for this example is adding $1ms$ from the moment that the input is changed to the cell's response, to realize the GoL rules in a synchronous matter and avoid the loss of signals. This procedure is inherent to the MFC configuration as the transition time between states is reported to be consistent around four minutes \cite{greenman2006microbial}. 

\section*{Conclusions}
The capabilities of MFCs in water purification, extraction of useful elements, sensory applications and energy production have been thoroughly studied. Another proposed application for MFCs is the carrying out of computational functions, which has been expressed as the construction of conventional logic gates, a Pavlovian learning model and, in this study, an implementation of a CA paradigm, namely GoL. 

The ability of interconnecting MFCs via hydraulic and electrical links and the multiple states that can be adopted by each MFC, makes the possible computing configurations more complex than conventional ones. Moreover, the MFC computing units are not limited by a power source; on the contrary they are powered by sustainable, diverse and abundant fuel sources. A disadvantage of these systems is the long transition times experienced between steady states that can reach up to four minutes; however, they have been reported to be consistent. 

Here the design of a duet of MFCs interconnected hydraulically and electrically to form a unit that behaves like a cell of GoL was proposed. Namely, the effluent of one MFC is used as an influent of the other, the third electrodes of both are connected with the electrical input of the cell, while the anode of one of the MFCs is used as the output of the cell. Given the fact that the proposed configuration consisted of two MFCs has the same behavior as a GoL CA cell, the realization of universal computation is possible.

An aspect of future work is the implementation of different CA local rules with configurations of real interconnected MFCs. Furthermore, the possibility of using conventional computing machines for the initialization of the CA grid and the exploitation of the outputs will be investigated, to design and realize a hybrid computing system.

\section*{Acknowledgments}
This work was supported by the European Union's Horizon 2020 Research and Innovation Programme under Grant Agreement No. 686585.

\nolinenumbers

%
%

\end{document}